\documentclass[a4paper,11pt]{article}
\usepackage{pos}

\newcommand\beq{ \begin{eqnarray} }
\newcommand\eeq{ \end{eqnarray} }

\title{Phase and equation of state of finite density QC$_2$D at lower temperature}

\author*[a,b]{Etsuko Itou}
\author[c,d]{Kei Iida}
\author[b,e]{Kotaro~Murakami,}
\author[f,g]{Daiki~Suenaga}

\affiliation[a]{Yukawa Institute for Theoretical Physics, Kyoto University, Kitashirakawa Oiwakecho, Sakyo-ku, Kyoto 606-8502, Japan }
\affiliation[b]{Interdisciplinary Theoretical and Mathematical Sciences Program (iTHEMS), RIKEN, Wako, Saitama 351-0198, Japan}  

\affiliation[c]{Department of Mathematics and Physics, Kochi University, 2-5-1 Akebono-cho, Kochi 780-8520, Japan }
\affiliation[d]{RIKEN Nishina Center, Wako, Saitama 351-0198, Japan}

\affiliation[e]{Department of Physics, Institute of Science Tokyo, 2-12-1 Ookayama, Megro, Tokyo 152-8551, Japan }

\affiliation[f]{Kobayashi-Maskawa Institute for the Origin of Particles and the Universe, Nagoya University, Nagoya, 464-8602, Japan}
\affiliation[g]{Research Center for Nuclear Physics,
Osaka University, Ibaraki 567-0048, Japan }

\emailAdd{itou@yukawa.kyoto-u.ac.jp}

\abstract{We investigate the phase structure and the equation of state (EoS) for dense two-color QCD at low temperatures, $T = 40$ MeV ($32^4$ lattice) and $T = 80$ MeV ($16^4$ lattice). A rich phase structure below the pseudo-critical temperature $T_c$ as a function of quark chemical potential $\mu$ has been revealed. By performing $T = 40$ MeV simulations, essentially similar results to the previous ones at $T = 80$ MeV are obtained, but several finer understandings are achieved. Breaking of the conformal bound is also confirmed thanks to smaller statistical errors.
This talk is mainly based on Refs.~\cite{Iida:2022hyy, Iida:2024irv}. It also includes related studies and subsequent developments that were not mentioned in the original papers.}

\FullConference{The 41st International Symposium on Lattice Field Theory (LATTICE2024)\\
 28 July - 3 August 2024\\
Liverpool, UK\\}

\begin{document}

\begin{flushright}
YITP-24-179, RIKEN-iTHEMS-Report-24
\end{flushright}

\maketitle

\section{Introduction}

The sign problem in three-color QCD at low temperatures and finite densities still remains one of the most challenging issues in lattice Monte Carlo simulations~\cite{Nagata:2021ugx}. Meanwhile, multiple groups have actively conducted large-scale first-principles calculations using the (R)HMC algorithm for systems such as two-color QCD with baryon chemical potential and three-color QCD with isospin chemical potential, where the sign problem is absent. Adding explicit symmetry-breaking terms for the U(1) baryon or isospin symmetry (i.e., diquark or pionic source terms) into the finite-density QCD(-like) action has enabled the accumulation of gauge-field configurations in the superfluid phase, which yields various observables that are clearly different from those in the hadron phase and the quark-gluon plasma (QGP) phase.

Most remarkably, first-principles calculations for these QCD-like theories have attracted much attention in the context of breaking the conformal (holographic) bound~\cite{Cherman:2009tw, Hohler:2009tv}; the speed of sound in the superfluid phase exceeds the value of relativistic free theory corresponding to $c_{\mathrm s}^2/c^2 = 1/3$.
The first evidence for such breaking was obtained in dense two-color QCD~\cite{Iida:2022hyy, Iida:2024irv}; shortly afterward, it was independently confirmed in the case of three-color QCD with isospin chemical potential using two different computational approaches~\cite{Brandt:2022hwy, Abbott:2023coj, Abbott:2024vhj}~\footnote{Also, at the conference, an independent analysis of the breaking of the conformal bound for dense two-color QCD was reported~\cite{Hands:2024nkx}. }.

In this article, we summarize our results for the phase diagram and equation of state (EoS) of two-color QCD at low temperatures and finite densities~\cite{Iida:2019rah, Ishiguro:2021yxr, Iida:2022hyy,  Iida:2024irv}. Furthermore, we will compare our results for the speed of sound and EoS with related lattice calculations, effective model studies, and observational data on neutron stars. 
In particular, we focus on related works that remain to be cited in our previous paper~\cite{Iida:2024irv}.

\section{Phase diagram}
For the lattice action, we employed the Iwasaki gauge action and naive Wilson fermions. Following the approach described in Refs.~\cite{Iida:2019rah, Iida:2022hyy, Iida:2024irv}, we extended the two-color QCD action by including a number operator coupled with the quark chemical potential ($\mu$). 
In low-temperature and high-density regimes, furthermore, we introduced a diquark source term characterized by a parameter $j$;  physical quantities  were calculated for several finite values of $j$, and then obtained by taking the $j\rightarrow 0$ limit.  Using the above-mentioned action,  we performed the RHMC calculations to generate gauge configurations at $\beta = 0.80$, $\kappa = 0.159$ for $32^4$ ($T = 40$ MeV), $16^4$ ($T = 80$ MeV), and  $32^3 \times 8$ ($T = 160$ MeV). The chemical potential ($\mu$) is normalized by the pseudo-scalar meson (corresponding to a pion in QCD) mass at $\mu=0$, namely $\mu/m_{\rm PS}$, where $m_{\rm PS}\approx 738$ MeV in our simulations~\cite{Murakami:2022lmq, Murakami:2023ejc}.

\begin{figure}[htbp]
\centering
\includegraphics[width=0.7 \textwidth]{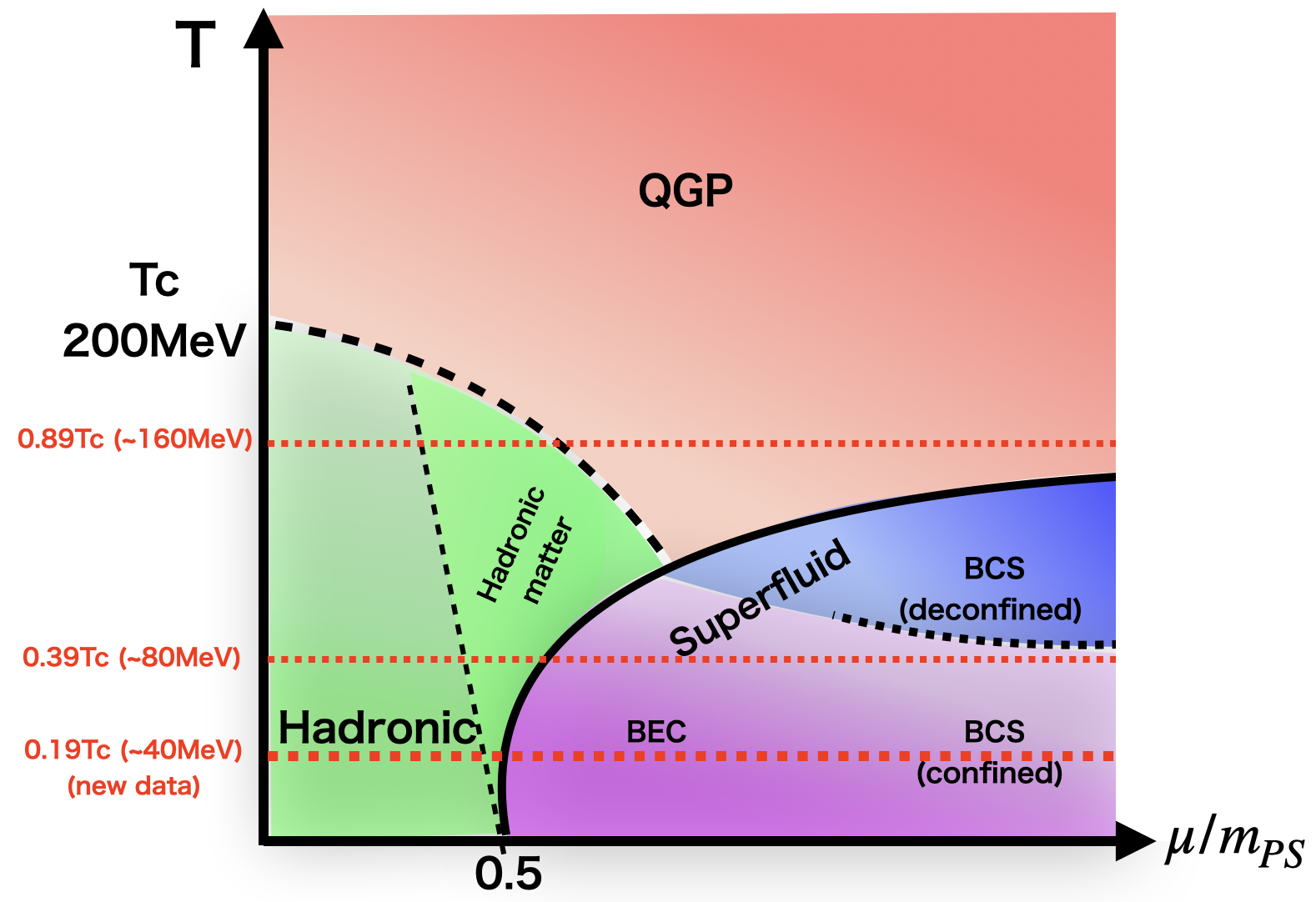}
\caption{Schematic QC$_2$D phase diagram. In our previous work~\cite{Iida:2019rah}, we clarified the phase structure at $T = 160$ MeV and 80 MeV, while in our recent work, we addressed what it is like at $T = 40$ MeV.}\label{fig:schematic}
\end{figure}

\begin{table}[h]
\begin{center}
\begin{tabular}{|c||c|c|c|c|c|}
\hline
 \multicolumn{1}{|c||}{}  & \multicolumn{2}{c|}{Hadronic } &     \multicolumn{2}{c|}{Superfluid } &\multicolumn{1}{c|}{QGP} \\  
\cline{3-3} \cline{4-5}  & & Hadronic matter ($T>0$) &BEC & BCS  &  \\  
 \hline \hline
$\langle |L| \rangle$ & zero  & zero  &   &   & non-zero \\
$\langle qq \rangle$ & zero  &  zero  & non-zero & ($\propto \mu^2$)  & zero \\ 
$ \langle n_q \rangle $ &   zero &  non-zero  & non-zero & $\langle n_q \rangle /n_q^{\rm tree} \approx 1$ & non-zero \\ 
 \hline
\end{tabular}
\caption{ Definition of the phases. To distinguish between the BEC and BCS phases, we use the value of $\langle n_q \rangle$. Meanwhile, $\langle qq \rangle$ is expected to scale as  $\propto\mu^2$ by the weak coupling analysis~\cite{Schafer:1999fe, Hanada:2011ju, Kanazawa:2013crb}. } \label{table:phase}
\end{center}
\end{table}
First, we obtained the phase diagram on $T$--$\mu$ plane as shown in Fig.~\ref{fig:schematic}.
Here, we utilize three quantities,  namely, the magnitude of the Polyakov loop ($\langle |L| \rangle$), the diquark condensate ($\langle qq \rangle$), and the quark number density ($\langle n_q \rangle$) to distinguish among different phases.
We use the name of each phase as shown in Table~\ref{table:phase}, which follows Refs.~\cite{Iida:2019rah,Hands:2006ve, Braguta:2016cpw, Boz:2019enj}.

Let us focus on the two relatively low temperatures, $T = 40$ MeV and $T = 80$ MeV\@.
At both temperatures, the diquark condensate, which is an order parameter of the superfluidity, becomes non-zero beyond $\mu \approx m_{\rm PS}/2$. The chiral perturbation theory (ChPT) gives the critical chemical potential as $\mu_c = m_{\rm PS}/2$ and the scaling behavior around $\mu_c$ as
\beq
\langle qq \rangle =A (\mu -\mu_c)^{1/2}.\label{eq:qq-ChPT-scaling}
\eeq
Our data are almost consistent with these predictions. Indeed, we obtain the critical value at $T = 40$ MeV from the fitting of the data as
$\mu/m_{\rm PS}=0.47$.
At $T = 80$ MeV,  the ``hadronic-matter phase'' was clearly observed, where $\langle qq \rangle =0 $ but $\langle n_q \rangle \ne 0$  as defined in Table \ref{table:phase}.  Interestingly, the study at $T = 40$ MeV revealed that 
this phase shrinks with respect to $\mu$ at lower temperature. It indicates that $\langle n_q \rangle \ne 0$ before the superfluid phase transition is caused by a thermal excitation of diquarks, which  correspond to the lightest hadron in the superfluid phase, at finite temperature.

Another notable finding from the comparison of the two temperatures concerns the scaling behavior of $\langle qq \rangle$ in the BCS phase. The scaling looks nearly linear in $\mu$ at $T = 80$ MeV, whereas a quadratic scaling was observed at $T = 40$ MeV\@. Analytical studies at zero temperature based on the weak-coupling expansion predict $\langle qq \rangle \propto \mu^2$~\cite{Schafer:1999fe, Hanada:2011ju, Kanazawa:2013crb}. Our results show that this analytical prediction does not hold at higher temperatures, but become closer to this prediction at lower temperatures. Also, in general, $\langle qq\rangle $ can be expressed approximately as $\Delta (\mu) \mu^2$, where $\Delta (\mu)$ is the diquark gap. The fact that $\langle qq\rangle $ behaves as $\mu^2$ suggests that the $\mu$-dependence of the diquark gap in the BCS phase is small.

\begin{figure}[htbp]
\centering
\includegraphics[width=1.0 \textwidth]{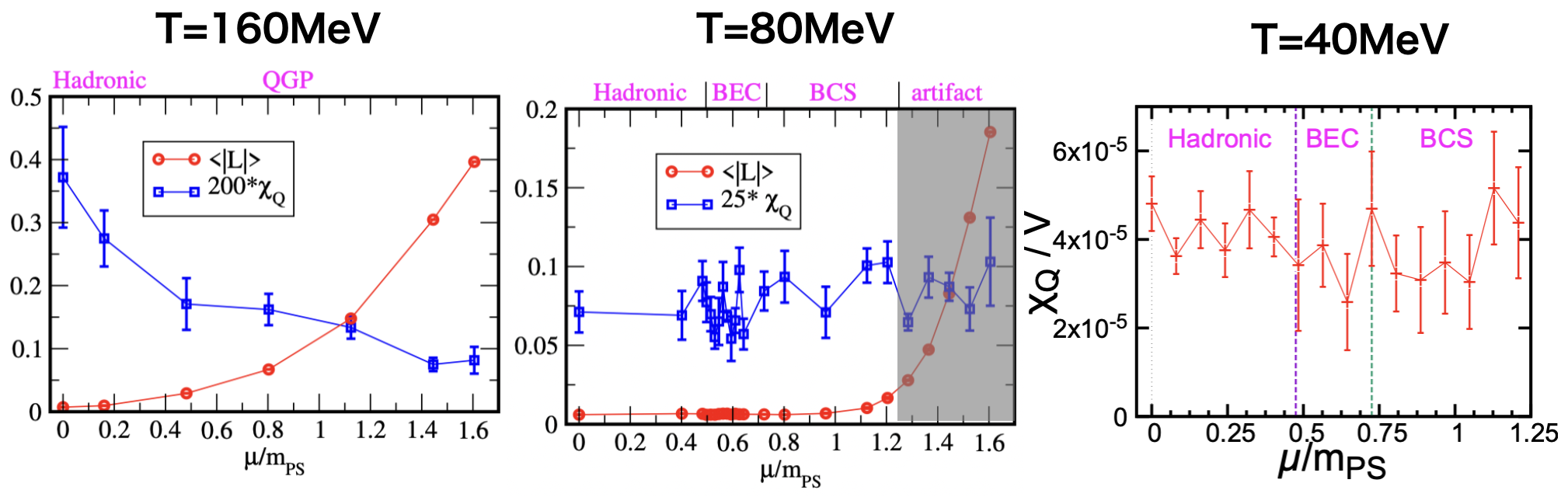}
\caption{The $\mu$-dependence of the topological susceptibility (red data) at $T = 160$ MeV, 80 MeV, and 40 MeV\@. At $T = 160$ MeV and 80 MeV, we also show the magnitude of the Polyakov loop (blue data) to see the confining behavior.}\label{fig:chi-Q-old}
\end{figure}
After determining the phase diagram, we investigated the $\mu$-dependence of the topological susceptibility in Refs.~\cite{Iida:2024irv, Iida:2019rah},
\beq
\chi_Q = \langle Q^2 \rangle - \langle Q \rangle^2.
\eeq
Here, $Q$ denotes the topological charge, which we  measured from the gluonic definition,
\beq
Q (t) = \frac{1}{32 \pi^2} \sum_{x} \mbox{Tr} \epsilon_{\mu \nu \rho \sigma} 
G^a_{\mu \nu} (x,t) G^a_{\rho \sigma} (x,t),
\eeq
using the gradient flow method.
At $T = 160$ MeV, where the hadronic to QGP phase transition occurs as $\mu$ increases, $\chi_Q$ decreases with $\mu$ as shown in the left panel of Fig.~\ref{fig:chi-Q-old}. On the other hand, at low temperatures of $T = 80$ MeV (the middle panel) and 40 MeV (the right panel), where the hadron-superfluid transition occurs, $\chi_Q$ remains almost constant throughout the $\mu$ region that covers the hadronic and superfluid phases. We also showed that the confinement remains even in the BCS phase, by studying the Polyakov loop (as shown in the blue data of the middle panel) and also $q$--$\bar{q}$ potential at $T = 40$ MeV in Ref.~\cite{Ishiguro:2021yxr}.
Although the high-density region is naively expected to be approximated by an asymptotically free theory, the macroscopic gluonic dynamics behaves similarly to  that in the hadronic phase.

\section{Equation of state}
Next, we consider the EoS at $T = 40$ MeV and $T = 80$ MeV\@. We calculated the trace anomaly and pressure using the following definitions.  
As for the pressure, we  employed
\beq
\frac{p}{p_{SB}}(\mu) &= \frac{\int_{\mu_o}^{\mu} d\mu' \frac{n_{SB}^{cont.}}{n_{q}^{\mathrm{tree}}}   n^{latt.}_q(\mu')  }{\int_{\mu_o}^{\mu} d\mu' n_{SB}^{cont.} (\mu')},\label{eq:p-scheme2}
\eeq
which was originally proposed in Ref.~\cite{Hands:2006ve}.
Here, $p_{SB}(\mu)$,  i.e., the pressure value of the non-interacting theory (the Stefan-Boltzmann (SB) limit),  was obtained by the numerical integration of the number density of  massless quarks.
As for the trace anomaly, we  utilized
\beq
e-3p &=& \frac{1}{N_s^3 N_\tau} \left( \left. a \frac{d \beta}{d a} \right|_{\mathrm{LCP}} \left\langle \frac{\partial S}{\partial \beta} \right\rangle_{sub.}  + \left. a \frac{d  \kappa}{d a} \right|_{\mathrm{LCP}} \left\langle \frac{\partial S}{\partial \kappa} \right\rangle_{sub.} + \left. a \frac{\partial j}{\partial a}\right|_{\mathrm{LCP}} \left\langle \frac{\partial S}{\partial j} \right\rangle_{sub.} \right), \nonumber \\  \label{eq:trace-anomaly}
\eeq
where the nonperturbative $\beta$-function was calculated in Ref.~\cite{Iida:2020emi}  as
\beq
 a d\beta /da|_{\beta=0.80,\kappa=0.159}=-0.352, \quad a d\kappa/da |_{\beta=0.80,\kappa=0.159}=0.0282.\label{eq:beta-fn}
\eeq

\begin{figure}[htbp]
\centering
\includegraphics[width=.4\textwidth]{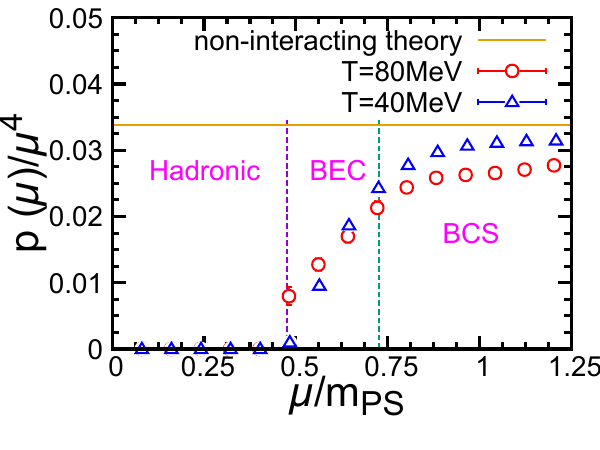}
\qquad
\includegraphics[width=.4\textwidth]{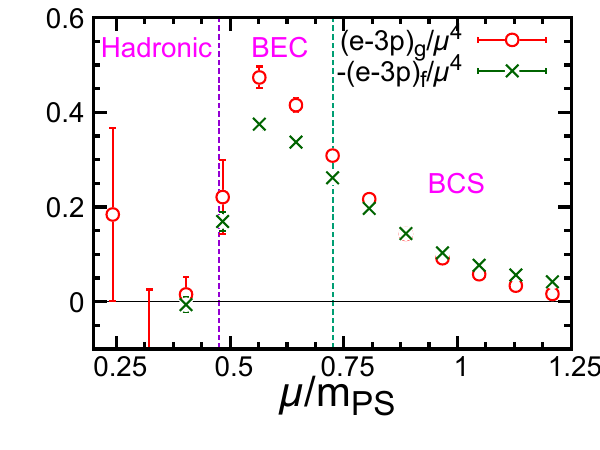}
\caption{Pressure (left panel) and trace anomaly (right panel) as a function of $\mu$.   The right panel  depicts the first (circle-red symbol) and second (cross-blue symbol) terms in Eq.~\eqref{eq:trace-anomaly} at $T = 40$ MeV separately. }\label{fig:p-Tdeps}
\end{figure}

The results for the pressure are shown in the left panel of Fig.~\ref{fig:p-Tdeps}. In the hadronic phase, where $\langle n_q \rangle $ is consistent with zero, the pressure is also zero. Once the superfluid phase transition occurs, the pressure increases monotonically. At $T = 40$ MeV (triangle-blue symbols), the pressure grows sharply in the BEC phase and approaches the SB limit more closely in the high-density region.  
On the other hand, we plotted the first (gluonic) term and minus the second (fermionic) term of the trace anomaly~\eqref{eq:trace-anomaly} at $T = 40$ MeV as $\langle e- 3p\rangle_g$ (circle-red symbols) and  $- \langle e -3p \rangle_f$ (cross-green symbols), respectively, in the right panel of Fig.~\ref{fig:p-Tdeps}. Note that the second term takes negative values, which have the sign flipped in the plot. Furthermore, we neglected the third term in Eq.~\eqref{eq:trace-anomaly} in our analysis.
Thus, the total trace anomaly is given by the circle-red data minus the cross-green data.
As can be seen from this plot, the trace anomaly is also zero in the hadronic phase. After the superfluid phase transition, the trace anomaly reaches a maximum value and then decreases. Notably, in the middle of the BCS phase, $- \langle e - 3p \rangle_f$ becomes larger than $\langle e -3p\rangle_g$, causing the trace anomaly to change from positive to negative.  

We would like to discuss the results that were obtained thus far.  
First, let us consider why the pressure value in the BCS phase at $T = 40$ MeV is larger than at $T= 80$ MeV\@. One would expect that particles are thermally excited at higher temperatures, resulting in higher pressure. Our results, however, show the opposite. In fact, the spatial volumes differ between these simulations: we performed the simulation on lattices of size $32^4$ for $T = 40$ MeV and $16^4$ for $T = 80$ MeV\@. 
Thus, the data for $T =80$ MeV may be more severely affected by the finite volume effects. We consider that such  effects might lead to a smaller pressure in our data,  especially at $T = 80$ MeV\@.  As to how to define the pressure on the lattice in a finite-density regime,  furthermore, there is some room for discussion~\cite{Hands:2006ve}.

Additionally, in the thermodynamic limit  at zero temperature, there is a relationship between the trace anomaly and the $\mu$-derivative of the pressure~\footnote{E.~I. would like to thank Dr.~Yuki~Fujimoto for pointing out this discussion.},
\beq
\frac{d}{d\mu} \left( \frac{p}{\mu^4} \right) = \frac{e-3p}{\mu^5}.
\eeq
Our data are inconsistent with this equation since the trace anomaly changes its sign in the BCS phase while  $p/\mu^4$ increases monotonically. This also suggests that finite volume effects may influence our data at least in the BCS phase.

\begin{figure}[htbp]
\centering
\includegraphics[width=0.4 \textwidth]{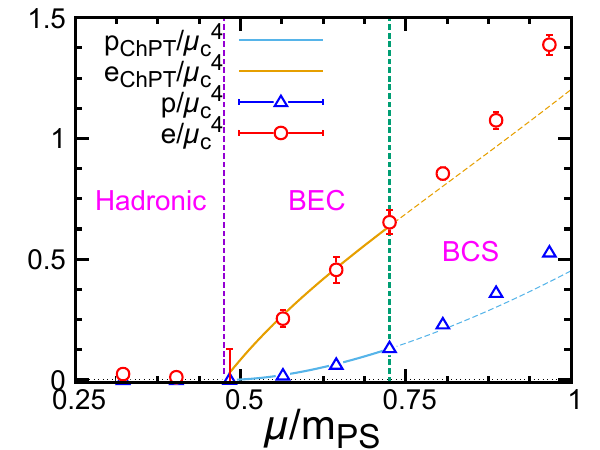}
\caption{ The pressure and internal energy around the BEC phase. The cyan and orange curves represent the fitting functions for $p/\mu_c^4$ and $e/\mu_c^4$, respectively, whose forms are given by the ChPT theory as shown in Eqs.~\eqref{eq:ChPT-p} and~\eqref{eq:ChPT-e}. }\label{fig:fpi}
\end{figure}
If we focus on the BEC phase, on the other hand, our results for the chiral condensate, diquark condensate, and sound velocity (to be shown later) are consistent with the predictions from ChPT~\cite{Iida:2024irv}.
It would therefore be worthwhile to fit the data for $p$ and $e$ in the BEC phase to the prediction of ChPT~\cite{Hands:2006ve},
\beq
p_{\rm ChPT}&=&4N_f F^2 \mu^2 \left( 1- \frac{\mu_c^2}{\mu^2} \right)^2, \label{eq:ChPT-p} \\
e_{\rm ChPT}&=&4N_f F^2 \mu^2 \left( 1- \frac{\mu_c^2}{\mu^2} \right) \left( 1+3 \frac{\mu_c^2}{\mu^2} \right), \label{eq:ChPT-e} 
\eeq
to obtain the pion decay constant ($F$) in two-color QCD.
Figure~\ref{fig:fpi} depicts the $\mu$ dependence of $p$ and $e$ both for the data and fitted results. 
The best fit values  were obtained as $F= 51.1(5)$ MeV and $F=56.7(7)$ MeV from the fits of $p/\mu_c^4$ and $e/\mu_c^4$, respectively.
These values are similar to the  earlier result, $F=60.8(1.6)$ MeV, obtained in Ref.~\cite{Astrakhantsev:2020tdl}
from the fit of  the lattice data for the quark number density and the mixing angle between the diquark and chiral condensates around the phase transition point predicted by the ChPT analysis.

\begin{figure}[htbp]
\centering
\includegraphics[width=0.45 \textwidth]{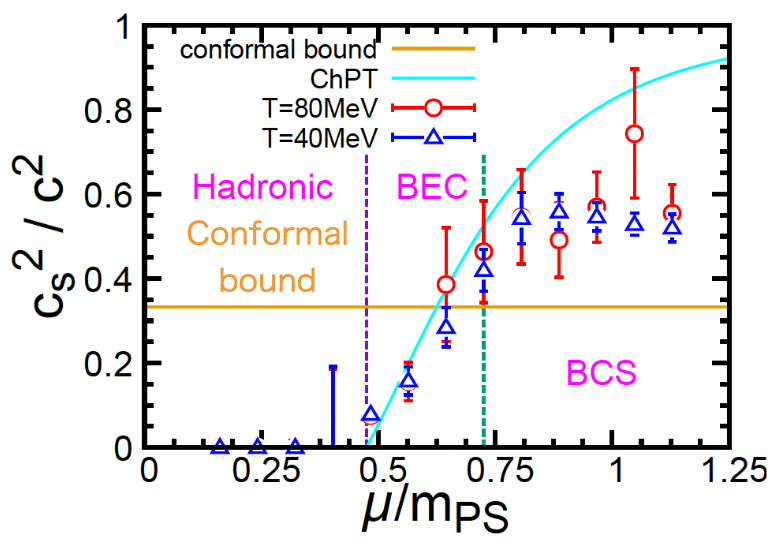}
\caption{The squared sound velocity at $T = 40$ MeV and $T = 80$ MeV. The cyan curve is the prediction of ChPT given by $c_{\rm s}^2/c^2=(1-\mu_c^4/\mu^4)/(1+3\mu_c^4/\mu^4)$. The horizontal line (orange) depicts the conformal bound, $c_{\rm s}^2/c^2 = 1/3$. }\label{fig:sound-v}
\end{figure}
Finally, we plotted the data for the speed of sound at $T = 40$ MeV and $T = 80$ MeV in Fig.~\ref{fig:sound-v}, where we evaluated
\beq
c_\mathrm{s}^2 (\mu)/c^2= \frac{\Delta p (\mu)}{\Delta e (\mu)} = \frac{ p(\mu +\Delta \mu) - p(\mu -\Delta \mu)}{e(\mu +\Delta \mu) - e(\mu -\Delta \mu)}\label{eq:sound-velocity}
\eeq
at fixed temperature.
The prediction of the ChPT,  which was shown as cyan-curve,  is that $c_{\mathrm s}^2/c^2 $ approaches $1$  with increasing density.  Our data are consistent with this prediction in the low-density region of the BEC phase.  In the high-density region  of the BEC phase, however, the values become smaller than those predicted by ChPT.
 While this behavior remains in the BCS phase,
the conformal bound (shown as orange-line), known as $c_{\mathrm s}^2/c^2 =1/3$, is  nevertheless clearly exceeded. Before our study~\cite{Iida:2022hyy}, for instance,  a number of finite temperature studies at $\mu=0$  had existed, but no first-principles calculation in any QCD-like theory had demonstrated a violation of this conformal bound. It is a characteristic property of low-temperature and high-density QCD-like theories,  which could help to support earlier suggestions of such a violation from neutron star phenomenology and effective model analyses~\cite{Masuda:2012ed,Baym:2017whm, McLerran:2018hbz,Fujimoto:2020tjc, Kojo:2021ugu, Kojo:2021hqh, Braun:2022jme}.

\section{Summary and related works}
We investigated the phase structure and the  EoS for dense two-color QCD at low temperatures: $T = 40$ MeV ($32^4$ lattice) and $T = 80$ MeV ($16^4$ lattice).  From a comparison of various observables between the two temperatures, we  found that the hadronic-matter phase shrinks as the temperature decreases and that the diquark condensate shows more of a quadratic scaling tendency in the BCS phase, which is predicted by the weak coupling expansion. Furthermore, careful analyses confirm that the topological susceptibility is independent of $\mu$ even in  a high-density regime. We also compared the data obtained at the two temperatures for the pressure, internal energy, and sound velocity as a function of $\mu$. The pressure increases around the hadronic-superfluid phase transition more rapidly at the lower temperature, while the temperature dependence of the speed of sound is invisible. Breaking of the conformal bound was also confirmed thanks to sufficiently small statistical errors at $T=40$ MeV\@.
It is interesting to note that recent lattice results for three-color QCD with isospin chemical potential $\mu_I$ confirm that the speed of sound agrees with ChPT in a relatively small $\mu_I$ regime  of the  pionic BEC phase and subsequently violates the conformal bound  at higher isospin densities~\cite{Brandt:2022hwy, Abbott:2023coj, Abbott:2024vhj}. 
As for the upper bound of the speed of sound, a recent paper~\cite{Hippert:2024hum} proposed a new value of the bound from a hydrodynamics analysis, which gives $c_{\mathrm s}^2/c^2 < 0.781$.
Our data also satisfy this bound.

It is worth mentioning several recent analytical studies that consider the origin of the large value of the speed of sound and the negativity of  the trace anomaly.
It has been pointed out that the presence of a pairing gap tends to increase the speed of sound compared with the perturbative analysis~\cite{Kojo:2014rca, Leonhardt:2019fua, Braun:2022jme}.
 According to a recent study by Fujimoto~\cite{Fujimoto:2024pcd}, the pressure  in three-color isospin QCD approaches the SB limit from above due to the gap. In contrast, in dense two-color QCD  where 
  corrections from the gap are relatively small, the pressure is only slightly modified from perturbation theory, and hence the pressure approaches the SB limit from below.   Such predictions look consistent with the high-density lattice results for three-color isospin QCD (Ref.~\cite{Abbott:2024vhj}) and for  dense two-color QCD shown here.   Furthermore, Fukushima and Minato have conducted studies on the negativity of the trace anomaly for  dense two-color QCD and three-color isospin QCD, as well as two-flavor color-superconducting (2SC) matter, using a unified treatment of  the perturbative gap and  the correction due to instanton effects, and thereby they have revealed differences among these systems~\cite{Fukushima:2024gmp}. In particular, 2SC matter shows a qualitatively different behavior for the speed of sound and the trace anomaly as a function of $\mu$ from the others. 
On the other hand, the constraints from neutron star observational data, which can be effectively regarded as those from real finite-density QCD, suggest that the conformal bound is violated~\cite{Altiparmak:2022bke, Annala:2023cwx}.
In conjunction with the results of these analytical  and phenomenological studies, it would be fascinating to delve into the similarities and differences among dense two-color QCD and three-color isospin systems, let alone, three-color QCD matter in our Universe.

\section*{Acknowledgment}
We would like to thank Drs.\ Y.~Fujimoto, K.~Fukushima,  S.~Hands, T.~Hatsuda, T.~Kojo, Y.~Namekawa, and N.~Yamamoto for helpful comments.
The work of K.~I. is supported by JSPS KAKENHI with Grant Numbers 18H05406 and 23K25864.
The work of E.~I. is supported by JST PRESTO Grant Number JPMJPR2113, 
JSPS Grant-in-Aid for Transformative Research Areas (A) JP21H05190, 
JST Grant Number JPMJPF2221,  
JPMJCR24I3,  
and also supported by Program for Promoting Researches on the Supercomputer ``Fugaku'' (Simulation for basic science: from fundamental laws of particles to creation of nuclei) and (Simulation for basic science: approaching the new quantum era), and Joint Institute for Computational Fundamental Science (JICFuS), Grant Number JPMXP1020230411. 
The work of E.~I. is supported also by Center for Gravitational Physics and Quantum
Information (CGPQI) at YITP.
E.~I and D.~S. are supported by JSPS KAKENHI with Grant Number 23H05439. 
K.~M. is supported in part by Grants-in-Aid for JSPS Fellows (Nos.\ JP22J14889, JP22KJ1870) and by JSPS KAKENHI with Grant No.\ 22H04917. 
The work of D.~S. is also supported by JSPS KAKENHI with Grant Number 23K03377.
The numerical simulation is supported by the HPCI-JHPCN System Research Project (Project ID: jh220021) and HOKUSAI in RIKEN.

\bibliographystyle{JHEP}
\bibliography{2color}

\end{document}